\documentclass{ws-mpla}

\newcommand{\nablas}{\setbox0\hbox{$\nabla$}%
    \rlap{\hbox to \wd0{\hss/\hss}}\box0}

\newcommand{\delder}[1]{\frac{\delta}{\delta #1}}   		

\newcommand{\unit}{\mathbf{1}}   			

\newcommand{\CA}{\mathcal{A}}    			

\newcommand{\xd}{\dot{x}}

\newcommand{\CG}{\mathcal{G}}

\newcommand{\CI}{\mathcal{I}}

\newcommand{\CL}{\mathcal{L}}

\newcommand{\CN}{\mathcal{N}}
\newcommand{\CO}{\mathcal{O}}

\newcommand{\CT}{\mathcal{T}}

\newcommand{\CE}{\mathcal{E}}
\newcommand{\frg}{\mathfrak{g}}				

\newcommand{\cA}{\mathring{A}}
\newcommand{\cF}{{\mathring{F}}}
\newcommand{\cX}{{\mathring{X}}}

\newcommand{\cPsi}{{\mathring{\Psi}}}

\newcommand{\FR}{\mathbb{R}}     			
\newcommand{\FC}{\mathbb{C}}     			
\newcommand{\RZ}{\mathbb{Z}}     			

\newcommand{\dd}{\mathrm{d}}     			
\newcommand{\dpar}{\partial}     			
\newcommand{\de}{\mathrm{e}}     			
\newcommand{\di}{\mathrm{i}}     			
\newcommand{\eps}{{\varepsilon}}			
\newcommand{\epsb}{{\bar{\varepsilon}}}			

\newcommand{\zb}{{\bar{z}}}
\newcommand{\Zb}{{\bar{Z}}}
\newcommand{\psib}{{\bar{\psi}}}

\newcommand{\eand}{{\qquad\mbox{and}\qquad}}     		
\newcommand{\ewith}{{\qquad\mbox{with}\qquad}}

\newcommand{\der}[1]{\frac{\dpar}{\dpar #1}}   		
\newcommand{\dder}[1]{\frac{\dd}{\dd #1}}   		
\newcommand{\tr}{\,\mathrm{tr}\,}     			
\newcommand{\nablabs}{\bar{\nablas}}

\newcommand{\au}{\mathfrak{u}}
\newcommand{\asu}{\mathfrak{su}}

\newcommand{\sU}{\mathsf{U}}     			

\newcommand{\sSU}{\mathsf{SU}}

\newcommand{\sSO}{\mathsf{SO}}

\newcommand{\sSpin}{\mathsf{Spin}}

\newcommand{\acton}{\vartriangleright}     			

\begin{document}

\markboth{Christian S\"amann}
{M-Brane Models and Loop Spaces}

\catchline{}{}{}{}{}

\title{M-Brane Models and Loop Spaces\footnote{Invited review to appear in MPLA based on a talk given at UC Berkeley on 31.5.2011.}}

\author{\footnotesize CHRISTIAN S\"AMANN}

\address{Department of Mathematics and \\
Maxwell Institute for Mathematical Sciences\\[0.1cm]
Heriot-Watt University\\
Colin Maclaurin Building, Riccarton\\
 Edinburgh EH14 4AS, U.K.\\[0.1cm]
c.saemann@hw.ac.uk
}

\maketitle

\pub{Received (Day Month Year)}{Revised (Day Month Year)}

\begin{abstract}
I review an extension of the ADHMN construction of monopoles to M-brane models. This extended construction gives a map from solutions to the Basu-Harvey equation to solutions to the self-dual string equation transgressed to loop space. Loop spaces appear in fact quite naturally in M-brane models. This is demonstrated by translating a recently proposed M5-brane model to loop space. Finally, I comment on some recent developments related to the loop space approach to M-brane models.
\begin{flushright}
\raisebox{11.04cm}[0cm]{HWM--12--07}\\
\raisebox{11.04cm}[0cm]{EMPG--12--10}
\end{flushright}
\vspace*{-1cm}
\end{abstract}

\ccode{PACS Nos.: 11.25.Yb; 02.30.Ik.}

\section{Introduction and results}

In recent years, more and more effort has been devoted to finding and studying field theories which are candidates for effective descriptions of multiple M-branes. Inspired by the proposal of Basu and Harvey\cite{Basu:2004ed} of an M-theory version of the Nahm equation, Bagger and Lambert\cite{Bagger:2007jr} and independently Gustavsson\cite{Gustavsson:2007vu} (BLG) found a three-dimensional Chern-Simons matter theory with $\CN=8$ supersymmetry that is invariant under parity transformations. This theory was based on a matter multiplet taking values in a 3-Lie algebra\cite{Filippov:1985aa} and a topological gauge field with a natural action on the matter fields.

It was soon realized that the number of 3-Lie algebras with desirable properties, such as admitting a positive definite metric, is very limited and generalizations of the BLG model were sought. The most promising such generalization is the Aharony-Bergman-Jafferis-Maldacena (ABJM) model\cite{Aharony:2008ug}, which is again a Chern-Simons matter theory. In this model the matter fields live in the bifundamental representation of the product gauge group $\sU(N)\times \sU(N)$, which is endowed with a metric of split signature. One can, however, rewrite the ABJM model in 3-algebra language if one uses a generalization of 3-Lie algebras known as hermitian 3-algebras\cite{Bagger:2008se}.

It seems that supersymmetry in the ABJM model is reduced to $\CN=6$, which makes it less appealing than the BLG model. This could be remedied by the inclusion of monopole operators, see \cite{Bagger:2012jb} and references therein. Furthermore, the ABJM model passed a non-trivial consistency check when the scaling $N^{3/2}$ of the degrees of freedom with the number $N$ of M2-branes was reproduced\cite{Drukker:2010nc}. For a comprehensive review of M2-brane models, see \cite{Bagger:2012jb}.

Given that there is now a reasonable candidate for the description of multiple M2-branes, it is clearly tempting to look for a similar candidate for the effective description of multiple M5-branes. There are various arguments in the literature suggesting that such a theory has to be intrinsically quantum and that it should not possess a Lagrangian description\cite{Witten:2007ct}. However, recent results suggest that there might be loopholes to these arguments, as e.g.\ the inclusion of non-local components in the model\cite{Chu:2012um}.

As mentioned above, the development of the M2-brane models used as a guideline the Basu-Harvey equation that captures the dynamics of self-dual strings, i.e.\ M2-branes ending on M5-branes. It is therefore reasonable to expect that the description of self-dual strings from the perspective of the M5-branes might shed light on potential M5-brane models. Moreover, self-dual strings are interesting by themselves, as they should lead to new integrable models with a wealth of non-trivial mathematical structures behind them.

In the following, we will review a description of self-dual strings using loop space\cite{Saemann:2010cp,Palmer:2011vx}. Loop space appears very naturally within M-theory, as one can see by the following na\"ive argument: Recall that the interactions of D-branes are governed by fundamental strings, whose endpoints induce an effective theory on the set of points in the D-branes' worldvolume. The interactions of M5-branes, however, are governed by M2-branes, whose boundaries form loops called self-dual strings in the M5-branes' worldvolume. It is therefore natural to expect an effective description on the loop space of the worldvolume. The use of loop space in M-theory has been advocated before\cite{Schreiber:2005ff,Gustavsson:2005aq,Gustavsson:2006ie}. Moreover, loop space equations for self-dual strings (which differ from the ones presented here) have been proposed\cite{Gustavsson:2008dy}. We will give more arguments in favor of a loop space description of M5-branes in the last section.

A configuration of (flat) M2-branes ending on M5-branes corresponds in string theory to D2-branes ending on D4-branes or D1-branes ending on D3-branes. The latter configuration has been identified\cite{Diaconescu:1996rk,Tsimpis:1998zh} as the string theory interpretation of the Atiyah-Drinfeld-Hitchin-Manin-Nahm (ADHMN) construction of monopoles\cite{Nahm:1981nb,Hitchin:1983ay}. The loop space description of self-dual strings that we present in the following is a rather literal lift of the original ADHMN construction to M-theory. There are quite a few interesting structures related to the ADHMN construction, such as twistor descriptions, Lax pairs, spectral curves, etc. Our M-theory lift of the ADHMN construction should therefore allow us to find the M-theory analogues of these structures. We will list some more recent results related to this point in the last section.

As a further motivation for loop space, we also review the reformulation\cite{Papageorgakis:2011xg} of a set of equations of motion that was proposed\cite{Lambert:2010wm} as an effective description of M5-branes on loop space. This reformulation leads to natural interpretations for some of the fields appearing in these equations as well as for some of the equations themselves. The BPS equations in this model essentially agree with the self-dual string equation on loop space.

This review is structured as follows. In section 2, we recall the definition of self-dual strings as well as their description from the perspective of the M2-branes in terms of the Basu-Harvey equation. Just as monopoles can be seen as characteristic classes of principal fiber bundles, self-dual strings yield characteristic classes of abelian gerbes, which we review in section 3. Section 4 discusses the lift of an abelian gerbe to a line bundle over loop space, which yields a loop space version of the self-dual string equation. In section 5, we demonstrate how to lift all the ingredients in the ADHMN construction of monopoles to M-theory. We also perform a few consistency checks and demonstrate that this lift reduces to the ADHMN construction in a very natural way. Section 6 gives a brief impression of how the construction of section 5 is extended to self-dual strings in the ABJM model. In section 7, we review the loop space interpretation of a proposal of M5-brane equations. We close in section 8 by summarizing the presented results and putting them into the context of some more recent findings related to M-theory.

\section{From monopoles to self-dual strings}

As stated in the introduction, it was the study of the BPS equation for multiple M2-branes\cite{Basu:2004ed} that ultimately led to the construction of the BLG model\cite{Bagger:2007jr,Gustavsson:2007vu}. Analogously, one might hope that studying the BPS equations for multiple M5-branes might provide crucial hints on their effective description. This motivates the study of self-dual strings\cite{Howe:1997ue} and their non-abelian extensions. In this section, we provide a concise review of the self-dual string soliton and its relation to monopoles.

We start from a D-brane configuration of $n$ D1-branes ending on a D3-brane. The worldvolumes of the branes extend in ten-dimensional flat target space of type IIB superstring theory as follows:
\begin{equation}\label{diag:D1D3}
\begin{tabular}{rcccccccc}
${\rm IIB}$& 0 & 1 & 2 & 3 & 4 & 5 & 6 & \ldots\\
D1 & $\times$ & & & & & & $\vdash$ \\
D3 & $\times$ & $\times$ & $\times$ & $\times$ & & & $0$
\end{tabular}
\end{equation}
We have coordinates $x^0,x^1,\ldots$, and, as usual, we mostly write $s$ instead of $x^6$. The symbol $\vdash$ indicates that the D1-branes end on the D3-brane located at $s=x^6=0$. In the following, we will be interested in the vacua of this configuration.

From the perspective of the D3-brane, the endpoints of the D1-branes appear like $n$ Dirac monopoles. The configuration is thus described by the {\em Bogomolny monopole equation} $F=\star \dd\phi$, where $F=\dd A$ is the curvature of the electromagnetic potential $A$ and $\phi$ is the Higgs field of the monopole configuration.

From the perspective of the D1-brane, however, the configuration \eqref{diag:D1D3} is described by the {\em Nahm equation}\cite{Diaconescu:1996rk,Tsimpis:1998zh}
\begin{equation}\label{eq:Nahm}
 \dder{s}X^i=\tfrac{1}{2}\eps^{ijk}[X^j,X^k]~.
\end{equation}
The $X^i$ are scalar fields taking values in the adjoint representation of $\au(n)$ and describe the fluctuations of the $n$ D1-branes in the directions parallel to the worldvolume of the D3-branes. Note that the configuration \eqref{diag:D1D3} is invariant under $\sSO(3)$ rotations in the $x^1,x^2,x^3$-directions, and this is reflected in the Nahm equation.

Both descriptions are linked by the Nahm transform\cite{Nahm:1981nb,Hitchin:1983ay}. One can see this link very explicitly when comparing solutions to the Bogomolny monopole equation and the Nahm equation. Note that $F=\star \dd \phi$, together with the Bianchi identity, implies that $\phi$ is a harmonic function on $\FR^3$, where we allow for sources at the positions of the monopoles. For a monopole at $r:=|x|=0$, we therefore have a Higgs field $\phi\sim \frac{n}{r}$, where $n$ is the {\em monopole charge} and the number of D1-branes. On the other hand, we find a simple solution to the Nahm equation by factorizing $X^i=r(s)G^i$, where $G^i\in \au(n)$. Plugging this ansatz into \eqref{eq:Nahm}, we obtain
\begin{equation}\label{eq:fuzzyfunnel}
 r(s)=-\frac{1}{s}\eand G^i=\tfrac{1}{2}\eps^{ijk}[G^j,G^k]~.
\end{equation}
This solution is called the {\em fuzzy funnel}\cite{Myers:1999ps}: Equations \eqref{eq:fuzzyfunnel} imply that the $G^i$ form a representation of $\asu(2)$, which suggests to interpret them as coordinates on a fuzzy sphere $S^2_F$. The radius of this sphere is given by the function $r(s)$. Thus, each point in the worldvolume of the D1-branes polarizes into a fuzzy sphere with a radius that diverges at $s=0$. Because the Higgs field $\phi$ describes fluctuations of the D3-brane's worldvolume in the $x^6=s$-direction (and therefore the coordinate $s$ should be identified with $\phi$), the profile of the D1-branes given by $r(s)$ agrees precisely with that given by the Higgs field from the D3-brane's perspective.

Let us now lift this configuration to M-theory. For this, we T-dualize along $x^5$ and interpret $x^4$ as the M-theory direction. We arrive at the configuration
\begin{equation}\label{diag:M2M5}
\begin{tabular}{rccccccc}
${\rm M}$ & 0 & 1 & 2 & 3 & 4 & 5 & 6 \\
M2 & $\times$ & & & & & $\times$ & $\vdash$ \\
M5 & $\times$ & $\times$ & $\times$ & $\times$ & $\times$ & $\times$ & $s_0$
\end{tabular}
\end{equation}
Here, the boundary of the M2-branes in the M5-brane's worldvolume looks like a string. Together with the self-duality of the three-form curvature in the M5-brane's worldvolume, this led to the name {\em self-dual string}. We will assume that the boundary of the M2-brane is stable in the $x^5$-direction, and therefore restrict our attention to the $x^1,x^2,x^3,x^4$-directions. From the perspective of the M5-brane, the vacua of the configuration \eqref{diag:M2M5} are described by the self-dual string equation\cite{Howe:1997ue} $H:=\dd B=\star \dd \phi$.

The configuration \eqref{diag:M2M5} is obviously invariant under $\sSO(4)$ rotations in the $x^1,x^2,x^3,x^4$-directions, and this should be reflected in the extension of the Nahm equation. Basu and Harvey therefore suggested essentially the following equation\cite{Basu:2004ed}
\begin{equation}\label{eq:BasuHarvey}
 \dder{s}X^\mu=\tfrac{1}{3!}\eps^{\mu\nu\kappa\lambda}[X^\nu,X^\kappa,X^\lambda]~.
\end{equation}
Here, we assume that the bracket $[\cdot,\cdot,\cdot]$ is totally antisymmetric and linear in each slot. Before discussing this 3-algebra structure in any further detail, let us briefly compare the profiles of the scalar fields in the two descriptions. We make again a product ansatz $X^\mu=r(s)G^\mu$ and plug it into \eqref{eq:BasuHarvey}. The solution is
\begin{equation}\label{eq:fuzzy3funnel}
 r(s)=\frac{1}{\sqrt{2s}}\eand
 G^\mu=-\tfrac{1}{3!}\eps^{\mu\nu\kappa\lambda}[G^\nu,G^\kappa,G^\lambda]~.
\end{equation}
On the other hand, the self-dual string equation $H=\dd B=\star \dd \phi$ implies that $\phi$ is a harmonic function on $\FR^4$ with sources at the location of the self-dual string. We therefore have $\phi\sim \frac{n}{2r^2}$, which is again compatible with the profile obtained from the solution given in \eqref{eq:fuzzy3funnel}.

The components $G^\mu$ of the solution $X^\mu$ take values in a {\em 3-Lie algebra}\cite{Filippov:1985aa}: They are elements in a vector space $\CA$ endowed with a totally antisymmetric trilinear bracket $[\cdot,\cdot,\cdot]:\CA^{\wedge 3}\rightarrow \CA$. We also impose the {\em fundamental identity},
\begin{equation}\label{eq:fundamentalIdentity}
 [a,b,[c,d,e]]=[[a,b,c],d,e]+[c,[a,b,d],e]+[c,d,[a,b,e]]~,
\end{equation}
which plays the role of a higher Jacobi identity. It guarantees that the inner derivations $\frg_\CA$ spanned by the operators $D(a,b)$ with $D(a,b)\acton c:=[a,b,c]$, $a,b,c\in\CA$, form a Lie algebra. This Lie algebra can be used as a gauge algebra in the Basu-Harvey equation: Analogously to the Nahm equation, we can write \eqref{eq:BasuHarvey} in gauge covariant form after introducing a gauge potential $A_s$ taking values in $\frg_\CA$:
\begin{equation}\label{eq:BasuHarveyCovariant}
 \nabla_sX^\mu:=\dder{s}X^\mu+A_s\acton X^\mu=\tfrac{1}{3!}\eps^{\mu\nu\kappa\lambda}[X^\nu,X^\kappa,X^\lambda]~.
\end{equation}

The most prominent example of a 3-Lie algebra is the 3-Lie algebra $A_4\cong \FR^4$ whose generators $\tau_\mu$, $\mu=1,\ldots,4$, satisfy the relation
\begin{equation}
 [\tau_\mu,\tau_\nu,\tau_\kappa]=\eps_{\mu\nu\kappa\lambda}\tau_\lambda~.
\end{equation}
It underlies the solution \eqref{eq:fuzzy3funnel}, and its Lie algebra of inner derivations $\frg_{A_4}$ is $\sSO(4)$.

The Basu-Harvey equation is a BPS equation in the BLG model\cite{Bagger:2007jr,Gustavsson:2007vu}, a Chern-Simons matter theory with $\CN=8$ supersymmetry that was proposed as a candidate for the effective description of stacks of multiple M2-branes. In this theory, the matter supermultiplet, which takes values in a 3-Lie algebra, is coupled to a Chern-Simons action. To write down an action for this theory, it is necessary to equip the 3-Lie algebra with a metric structure. Unfortunately, $A_4$ and its direct sums are the only examples of finite-dimensional 3-Lie algebras that can be endowed with a positive definite, $\frg_\CA$-invariant metric. Generalizations of 3-Lie algebras were therefore proposed that relax the total antisymmetry of the 3-bracket\cite{Bagger:2008se,Cherkis:2008qr}. {\em Real 3-algebras}\cite{Cherkis:2008qr} are a generalization of 3-Lie algebras obtained by allowing the 3-bracket to be antisymmetric only in the first two slots. They evidently contain 3-Lie algebras. {\em
Hermitian 3-algebras}\cite{Bagger:2008se} are complex vector spaces endowed with a bracket $[\cdot,\cdot;\cdot]$, which is linear and antisymmetric in its first two slots and antilinear in the last one. It satisfies the following form of the fundamental identity:
\begin{equation}\label{eq:h3fundamentalIdentity}
 [[a,b;c],d;e]=[[a,d;e],b;c]+[a,[b,d;e];c]-[a,b;[c,e;d]]~.
\end{equation}
The 3-Lie algebra $A_4$ is contained in this class.

Both for real and hermitian 3-algebras, we have again a Lie algebra of inner derivations and we can construct Chern-Simons matter theories using them. In the case of hermitian 3-algebras, these theories contain the ABJM model\cite{Aharony:2008ug}, that is now regarded as the most promising candidate for the description of multiple M2-branes. More details on 3-algebras can be found in \cite{deAzcarraga:2010mr}.

All our constructions in the following can be applied to models using real or hermitian 3-algebras.

\section{Abelian gerbes}

While monopoles are described in terms of certain connections on principal fiber bundles, self-dual strings are described by certain connective structures on (abelian) gerbes. To explain this statement, let us briefly recall the bundle description of magnetic monopoles. For simplicity, we restrict ourselves to the abelian case.

Consider a manifold $M$ covered by patches $(U_i)$. Let $\CE$ be a circle- or principal $\sU(1)$-bundle over $M$. The {\em first Chern class} of $\CE$ can be identified with the curvature $F$ of a connection $\nabla$. Here, $F$ is a globally defined closed two-form with values in $\au(1)$. The Poincar\'e lemma allows us to derive gauge potentials $A_{(i)}$ on the patches $U_i$ with $F|_{U_i}=\dd A_{(i)}$. Another application of the Poincar\'e lemma yields transition functions $g_{(ij)}$ on the intersections of patches $U_i\cap U_j$ taking values in the group $\sU(1)$. Altogether we have
\begin{equation}
\begin{aligned}
F\in \Omega^2(M,\au(1))&~,\\
A_{(i)}\in \Omega^1(U_i,\au(1))&~\mbox{with}~F=\dd A_{(i)}~,\\
g_{(ij)}\in \Omega^0(U_i\cap U_j,\sU(1))&~\mbox{with}~A_{(i)}-A_{(j)}=\dd \log g_{(ij)}~.
\end{aligned}
\end{equation}
An example of a principal $\sU(1)$-bundle over $S^2$ is the Hopf fibration $\sU(1)\rightarrow S^3\rightarrow S^2$.

Consider now a Dirac monopole in $\FR^3$. As we have seen before, the Higgs field becomes singular at the position of the monopole. Let us therefore restrict to the unit sphere $S^2$ with the monopole at its center. We cover $S^2$ with the standard patches $U_+$ and $U_-$ centered at the north and south poles, respectively. The overlap $U_+\cap U_-$ can be contracted to the equator $S^1$. The transition function is therefore of the form $g_{(+-)}=\de^{-\di n\phi}$ with $n\in \RZ$ and $\phi\in[0,2\pi)$ the coordinate on the equator\footnote{For $n>0$, this bundle is therefore the $n$-fold tensor product of the principal $\sU(1)$-bundle given by the Hopf fibration. For $n<0$, we obtain tensor products of the dual of the Hopf fibration.}. The first Chern number of this principal $\sU(1)$-bundle is calculated as
\begin{equation}\label{eq:firstChern}
 c_1=\frac{\di}{2}\int_{S^2} F=\frac{\di}{2\pi}\int_{S^1} \left(A_{(+)}-A_{(-)}\right)=\frac{1}{2\pi}\int_0^{2\pi}\dd \phi~n =n~,
\end{equation}
where we used Stokes' theorem. The integer $n$ is a topological invariant and called the {\em monopole charge}.

Let us now come to the description of self-dual strings via abelian gerbes. Here, we will restrict ourselves to abelian local or Hitchin-Chatterjee gerbes, see e.g.\ \cite{Murray:2007ps,Hitchin:2010qz,Chatterjee:1998} for more details. Just as the first Chern class $F$ of a principal $\sU(1)$-bundle $\CE$ characterizes the bundle up to topological equivalence, the Dixmier-Douady class $H$ characterizes the abelian gerbe $\CG$. Here, $H$ is a globally defined, closed three-form with values in $\au(1)$. Repeated application of the Poincar\'e lemma yields the following objects:
\begin{equation}
 \begin{aligned}
H\in \Omega^3(M,\au(1))&~,\\
B_{(i)}\in \Omega^2(U_i,\au(1))&~\mbox{with}~H=\dd B_{(i)}~,\\
A_{(ij)}\in \Omega^1(U_i\cap U_j,\au(1))&~\mbox{with}~B_{(i)}-B_{(j)}=\dd A_{(ij)}~,\\
h_{(ijk)}\in \Omega^0(U_i\cap U_j\cap U_k,\au(1))&~\mbox{with}~A_{(ij)}-A_{(ik)}+A_{(jk)}=\dd h_{(ijk)}~.
\end{aligned}
\end{equation}
To describe a self-dual string, we restrict ourselves to a sphere $S^3$ with the self-dual string at its center. We cover $S^3$ by two patches $U_+$ and $U_-$, and the overlap $U_+\cap U_-$ can be contracted to the equator, which is a two-sphere. Instead of a transition function on the overlap, we have now a one-form potential of a connection and therefore a principal $\sU(1)$-bundle. As we saw above, however, the principal $\sU(1)$-bundles over $S^2$ are characterized by their monopole charge. Thus, abelian gerbes over $S^3$ are also characterized by an integer, which we call the {\em self-dual string charge}. Cohomologically, this is reflected in $F\in H^2(S^2,\RZ)\cong \RZ\cong H^3(S^3,\RZ)\ni H$. The self-dual string charge can be computed explicitly analogously to \eqref{eq:firstChern} by integrating $H$ over $S^3$.

\section{Self-dual strings on loop space}

As mentioned in the introduction, it is rather natural to expect a description of self-dual strings in terms of loop space. To perform the transition from abelian gerbes to objects on loop space, we use the so-called {\em transgression map}.

The free loop space of a manifold $M$ is the space of all maps $x:S^1\rightarrow M$. We restrict ourselves to maps that are immersions and factor out reparametrization invariance. The latter is done by working only with formulas that are invariant under reparametrizations. The resulting space of loops $\CL M$ is also called the space of singular oriented knots\cite{0817647309}. Acting on functions of $\CL M$ are the derivations
\begin{equation}
 X=\oint \dd \tau~X^{\mu \tau}~\delder{x^{\mu\tau}}:=\oint \dd \tau~X^{\mu}(\tau)~\delder{x^{\mu}(\tau)}~,
\end{equation}
where $x^\mu$ are some local coordinates on $M$. These derivations span the tangent space $T\CL M$. We write dual elements and the exterior derivative on $\CL M$ as
\begin{equation}
 \omega=\oint \dd \tau~\omega_{\mu \tau}~\delta x^{\mu\tau}~~~\eand~~~\delta=\oint \dd \tau~\delta x^{\mu\tau}~\delder{x^{\mu\tau}}~,
\end{equation}
respectively. For more details and the precise definition of the above objects, see \cite{0817647309}.

We have the following natural double fibration:
\begin{equation}
\begin{aligned}
\begin{picture}(50,40)
\put(0.0,0.0){\makebox(0,0)[c]{$M$}}
\put(64.0,0.0){\makebox(0,0)[c]{$\CL M$}}
\put(32.0,33.0){\makebox(0,0)[c]{$\CL M\times S^1$}}
\put(7.0,18.0){\makebox(0,0)[c]{$\mathsf{ev}$}}
\put(55.0,18.0){\makebox(0,0)[c]{$\mathsf{pr}$}}
\put(25.0,25.0){\vector(-1,-1){18}}
\put(37.0,25.0){\vector(1,-1){18}}
\end{picture}
\end{aligned}
\end{equation}
Here, $\mathsf{pr}$ is the obvious projection from a point $(x,\tau_0)\in \CL M\times S^1$ to $x\in \CL M$, and $\mathsf{ev}$ is the evaluation map $\mathsf{ev}(x,\tau_0)=x(\tau_0)$. Given a $k+1$-form $\omega$ on $M$, we can pull it back to $\CL M\times S^1$. Subsequent integration along the fibers of $\mathsf{pr}$ yields a $k$-form on $\CL M$. Locally, we have the following expressions:
\begin{equation}\label{eq:Transgression}
\begin{aligned}
 \omega=&\frac{1}{(k+1)!}\omega_{\mu_1\ldots\mu_{k+1}}(x)\dd x^{\mu_1}\wedge \cdots \wedge \dd x^{\mu_{k+1}} \mapsto\\& \CT \omega:=\oint \dd \tau~\frac{1}{k!}\omega_{\mu_1\ldots \mu_{k+1}}(x(\tau)) \xd^{\mu_{k+1}\tau}~\delta x^{\mu_1\tau}\wedge \cdots \wedge \delta x^{\mu_k\tau}~.
\end{aligned}
\end{equation}
The transgression map $\CT:\Omega^{k+1}(M)\rightarrow \Omega^k(\CL M)$ is clearly reparametrization invariant. Moreover, it is a chain map and maps closed forms to closed forms and exact forms to exact forms. It therefore descends to a map $\CT:H^{k+1}_{\rm dR}(M)\rightarrow H^k_{\rm dR}(\CL M)$ on de Rham cohomology. This implies that an abelian gerbe on $M$ characterized by a Dixmier-Douady class $H$ is mapped via transgression to a first Chern class $F$ of a principal $\sU(1)$-bundle over the loop space $\CL M$. Note, however, that $\CT$ is {\em not} surjective, and therefore there are in general more circle bundles over the loop space $\CL M$ than there are gerbes on $M$. Forms on loop space in the image of $\CT$ are usually called {\em ultra-local}.

Working with vector bundles instead of gerbes is a priori easier. For example, extending gauge structures to the non-abelian case is straightforward. It should be stressed, however, that in performing a transgression, we trade the difficulties of working with gerbes for the difficulties of working over infinite-dimensional spaces.

Let us now transgress the self-dual string equation $H=\star \dd \Phi$ on $\FR^4$ to the loop space $\CL \FR^4$. Applying $\CT$ to both sides of the equation, we obtain in Cartesian coordinates $x^\mu$, $\mu=1,\ldots,4$, on $\FR^4$ the following expression:
\begin{equation}\label{eq:preSDSequation}
 F_{(\mu\sigma)(\nu\rho)}=\delta(\sigma-\rho)\eps_{\mu\nu\kappa\lambda}\xd^\kappa(\tau) \left.\der{y^\lambda}\Phi(y)\right|_{y=x(\tau)}~.
\end{equation}
We would like to extend the above equation in various ways: First, $\Phi$ should be a function on $\CL M$ and not, as it stands, a function on $M$ or $\CL M\times S^1$. Moreover, we would like to extend the curvatures $F$ beyond abelian and ultra-local expressions. We will see that the appropriate generalization of \eqref{eq:preSDSequation} reads as\cite{Palmer:2011vx}
\begin{equation}\label{eq:SDSequation}
 \begin{aligned}
 F^\pm_{(\mu\sigma)(\nu\tau)}=&\big(\eps_{\mu\nu\kappa\lambda}\xd^\kappa(\sigma)\nabla_{(\lambda\tau)}\Phi\big)_{(\sigma\tau)}\\
&\hspace{0.5cm}\mp\big(\xd_{\mu}(\sigma)\nabla_{(\nu\tau)}\Phi+\xd_{\nu}(\sigma)\nabla_{(\mu\tau)}\Phi-\delta_{\mu\nu}\xd^\kappa(\sigma)\nabla_{(\kappa\tau)}\Phi\big)_{[\sigma\tau]}~,
\end{aligned}
\end{equation}
where
\begin{equation}
 \nabla_{(\mu\sigma)}:=\oint \dd \tau~ \delta x^\mu(\tau) \left(\delder{x^\mu(\tau)}+A_{(\mu\tau)}\right)
\end{equation}
and $[\sigma\tau]$ indicates antisymmetrization. We will refer to \eqref{eq:SDSequation} as the loop space self-dual string equation. Note that this equation is reparametrization invariant. In the following, we will demonstrate how to construct solutions to this equation from a generalized ADHMN construction.

\section{The generalized ADHMN construction}

The Atiyah-Drinfeld-Hitchin-Manin-Nahm (ADHMN) construction of monopoles is a Nahm transform, which is in turn a special Fourier-Mukai transform\cite{Schenk:1986xe}. The Nahm transform takes an instanton bundle over a four-torus $T^4$, pulls it back along the projection $\pi:T^4\times \hat{T}^4\rightarrow T^4$, where $\hat{T}^4$ is the four-torus dual to $T^4$, twists it with the Poincar\'e line bundle over $T^4\times \hat{T}^4$ and projects it down to an instanton bundle over $\hat{T}^4$. Roughly speaking, we consider in the ADHMN case a degenerate four-torus with radii $(0,0,0,\infty)$ yielding $\FR$. The ``dual four-torus'' has then radii $(\infty,\infty,\infty,0)$, giving $\FR^3$. The instanton equation reduces to the Nahm equation \eqref{eq:Nahm} on $\FR$ and the Bogomolny monopole equation $F=\star \dd\phi$ on $\FR^3$. In both cases, certain boundary conditions have to be imposed. The spaces $\FR$ and $\FR^3$ are to be seen as the spatial parts of the D1- and D3-branes in the configuration \eqref{diag:D1D3}.

Let us now go through the construction in more detail, and present all the ingredients that we will have to lift to M-theory. We start from a $\au(n)$-valued solution $X^i$ to the Nahm equation \eqref{eq:Nahm} on an open subset $\CI$ of $\FR$. For a Dirac monopole, we have $\CI=(0,\infty)$, while for an $\sSU(2)$-monopole, we choose $\CI=(-1,1)$. From this solution, we construct the {\em twisted Dirac operator}
\begin{equation}
 \nablas^{\rm IIB}_{s,x}=-\dder{s}+\sigma^i (\di X^i+x^i)\ewith \nablabs^{\rm IIB}_{s,x}:=\dder{s}+\sigma^i (\di X^i+x^i)~,
\end{equation}
which acts on $L_2$-sections of a trivial complex rank $n$ vector bundle tensored with the spinor bundle $S^+\cong \FC^2$ on $T^4$. The coordinates $x^i$ describing a point on $\FR^3$ arise from the twist of the vector bundle with the Poincar\'e line bundle. One can readily show that the $X^i$ satisfying the Nahm equation is equivalent to the Laplace operator $\Delta^{\rm IIB}_{s,x}:=\nablabs^{\rm IIB}_{s,x}\nablas^{\rm IIB}_{s,x}$ being positive and $[\Delta^{\rm IIB}_{s,x},\sigma^i]=0$.

We now determine all normalized zero modes of $\nablabs^{\rm IIB}_{s,x}$. That is, we find all $N$ solutions $\psi^a_{s,x}$, $a=1,\ldots,N$ such that
\begin{equation}
 \nablabs^{\rm IIB}_{s,x}\psi^a_{s,x}=0\eand\delta^{ab}=\int_\CI \dd s~\psib^a_{s,x}\psi^b_{s,x}~.
\end{equation}
A solution to the monopole equation on $\FR^3$ is then given by
\begin{equation}\label{eq:SolutionsFromNahm}
 A_i^{ab}:=\int_\CI \dd s~\psib^a_{s,x}\der{x^i}\psi^b_{s,x}\eand\Phi^{ab}:=-\di\int_\CI \dd s~\psib^a_{s,x}\,s\,\psi^b_{s,x}~,
\end{equation}
as one can verify by direct computation after introducing a Green's function for $\Delta^{\rm IIB}_{s,x}$. Let us briefly consider two examples for $n=1$ and $n=2$. For $n=1$, the Nahm equation reduces to $\dpar_sX^i=0$, and we can put $X^i=0$. A zero mode\footnote{There are in fact two zero modes $\psi_\pm$, yielding well-defined fields on the two domains\linebreak $\FR^3\backslash\{(0,0,x^3)|\pm x^3>0\}$ and corresponding to the Dirac string going through the north and the south pole, respectively.} is given by
\begin{equation}
 \psi_+=\de^{-s R}\frac{\sqrt{R+x^3}}{x^1-\di x^2}\left(\begin{array}{c}{x^1-\di x^2}\\{R-x^3}\end{array} \right)~,
\end{equation}
and using formulas \eqref{eq:SolutionsFromNahm}, we arrive at
\begin{equation}
 \Phi^+=-\frac{\di}{2R}~,~A^+_i=\frac{\di}{2(x^1+x^2)^2}\left(x^2\left(1-\frac{x^3}{R}\right),-x^1\left(1-\frac{x^3}{R}\right),0\right)~.
\end{equation}
In the case $n=2$, the Nahm equation is nontrivial, and we choose the solution we obtained before from factorization:
\begin{equation}
 X^i=-\frac{1}{s}G^i\ewith G^i=\frac{\sigma^i}{2\di}=-\bar{G}^i~.
\end{equation}
Following the ADHMN construction, one finds the solution
\begin{equation}
 \Phi^+=-\frac{\di}{R}~,
\end{equation}
and a more complicated expression for $A^+_i$. Recall that the Higgs field is expected to be proportional to the monopole charge $n$, which is indeed the case in the above examples.

To lift the construction up to M-theory, let us follow the T-duality along $x^5$ and the lift to M-theory along $x^4$ and construct Dirac operators at each step:
\begin{equation*}
\begin{tabular}{rcccccccl}
IIB & 0 & 1 & 2 & 3 & 4 & 5 & 6 \\
D1 & $\times$ & & & & & & $\times$ & $\nablas^{\rm IIB}_{s,x}=-\dfrac{\dd}{\dd s}+\sigma^i (\di X^i+x^i)$ \\
D3 & $\times$ & $\times$ & $\times$ & $\times$ & & \\[0.5cm]
IIA & 0 & 1 & 2 & 3 & 4 & 5 & 6 \\
D2 & $\times$ & & & & & $\times$ & $\times$ &$\nablas_{s,x}^{\rm IIA}=-\gamma_5\dfrac{\dd}{\dd s}+\gamma^4\gamma^i (X^i-\di x^i)$ \\
D4 & $\times$ & $\times$ & $\times$ & $\times$ & & $\times$ &\\[0.5cm]
M & 0 & 1 & 2 & 3 & 4 & 5 & 6 \\
M2 & $\times$ & & & & & $\times$ & $\times$ &$\nablas^{\rm M}_{s,x}=-\gamma_5\dfrac{\dd}{\dd s}+\tfrac{1}{2}\gamma^{\mu\nu}(D(X^\mu,X^\nu)-\di x^{\mu\nu})$ \\
M5 & $\times$ & $\times$ & $\times$ & $\times$ & $\times$ & $\times$ &
\end{tabular}
\end{equation*}
The Dirac operator $\nablas_{s,x}^{\rm IIA}$ is clear and has been introduced before\cite{Campos:2000de}. The $\sSO(4)$-invariant form of $\nablas^{\rm M}_{s,x}$ is equally evident, if we assume that the zero modes will take values in a 3-Lie algebra $\CA$. We can then contract $D(X^\mu,X^\nu)\acton c:=[X^\mu,X^\nu,c]$ with $\gamma^{\mu\nu}$. It is merely the twisting element $x^{\mu\nu}$ that is less clear. Here, however, the transition to loop space suggests to use
\begin{equation}
x^{\mu\nu}:=\oint \dd \tau\, x^{[\mu}(\tau)\xd^{\nu]}(\tau)~,
\end{equation}
where $x^\mu(\tau)$ is an element of $\CL \FR^4$, i.e.\ an equivalence class of maps $S^1\rightarrow \FR^4$. Altogether we obtain the twisted Dirac operator
\begin{equation}\label{eq:DiracM}
 \nablas^{\rm M}_{s,x}=-\gamma_5\dder{s}+\tfrac{1}{2}\gamma^{\mu\nu}\left(D(X^\mu,X^\nu)-\di \oint \dd \tau\, x^\mu(\tau)\xd^\nu(\tau)\right)~.
\end{equation}
This Dirac operator acts on $\CA$-valued functions on $\CI$ tensored by the trivial bundle $\FC^4$, potentially originating from the chiral spinors over $T^6$.

Let us now try to build a generalized ADHMN construction from this guess. First, one readily shows that for a solution $X^\mu$ to the Basu-Harvey equation \eqref{eq:BasuHarvey}, the Dirac operator \eqref{eq:DiracM} is positive and commutes with the generators $\gamma^{\mu\nu}$ of $\sSpin(4)$. 	The inverse statement is not true, as the map $D:\CA\wedge \CA\rightarrow \frg_\CA$ has a non-trivial kernel. Solutions for the loop space self-dual string equation \eqref{eq:SDSequation} are now obtained as follows: We start from the normalized zero modes $\psi^a_{s,x}$, $a=1,\ldots,2N$, satisfying
\begin{equation}\label{eq:normalizePsiM}
 \nablabs^{\rm M}_{s,x}\psi^a_{s,x}=0\eand\delta^{ab}=\int_\CI\dd s\, (\psib^a_{s,x},\psi^b_{s,x})~,
\end{equation}
where $(\cdot,\cdot)$ denotes the inner product on $\FC^4\otimes\CA$. Note that we have assumed here that $\CA$ is equipped with a compatible inner product $(\cdot,\cdot)$. Because of the block-diagonal structure of the Dirac operator \eqref{eq:DiracM} in a reasonable basis of the Clifford algebra, we can sort the zero modes according to their chirality. We arrive at $N$ zero modes $\psi^a_{s,x}$, $a=1,\ldots,N$, with $\gamma_5\psi^a_{s,x}=\psi^a_{s,x}$ and $N$ zero modes $\psi^a_{s,x}$, $a=N+1,\ldots,2N$, with $\gamma_5\psi^a_{s,x}=-\psi^a_{s,x}$. The solutions to \eqref{eq:SDSequation} are given by formulas very similar to \eqref{eq:SolutionsFromNahm}:
\begin{equation}\label{eq:MFieldDefinitions}
 A_{(\mu\tau)}^{ab}=\int \dd s\, \left(\psib^a_{s,x}, \delder{x^\mu(\tau)} \psi^b_{s,x}\right)\eand\Phi^{ab}=\di\int \dd s\, \left(\psib^a_{s,x},\,s\,\psi^b_{s,x}\right)~.
\end{equation}
One readily verifies that these fields satisfy the loop space self-dual string equation \eqref{eq:SDSequation} by explicit calculation. Note that they are anti-hermitian and, because of our sorting of zero modes, the fields take values in the gauge algebra $\au(N)_+\oplus\au(N)_-$.

To streamline the discussion, we ignored the role of boundary conditions in the ADHMN construction. There, one has to demand that the $X^i$ have simple poles $\frac{1}{s-s_0}$ at finite boundaries of $\CI$ and that they form irreducible representations of $\asu(2)$ at these points. There are corresponding boundary conditions in the M-brane case: The $X^\mu$ should behave like $\frac{1}{\sqrt{s-s_0}}$ at finite boundaries of $\CI$ and form irreducible representations of $\sSpin(4)$.

To explore our construction in more detail, let us first construct the obvious solutions for $n=1$ and $n=2$. In the case $n=1$, the Basu-Harvey equation becomes again $\der{s}X^\mu=0$ and we therefore put $X^\mu=0$. After introducing the shorthand $r_\pm^2:=\tfrac{1}{2}\sqrt{(x^{\mu\nu}\pm\tfrac{1}{2}\eps_{\mu\nu\kappa\lambda}x^{\kappa\lambda})^2}$, the zero modes are given by
\begin{equation}\label{eq:zmrSDk1N1}
\begin{aligned}
\psi^+_{s,x(\tau)}&\sim\de^{-r^2_- s}\left(\begin{array}{c}
\di \left(r_-^2+x^{12} -x^{34}\right) \\
x^{13}+x^{24}+\di (x^{23}-x^{14}) \\
0\\
0
\end{array}
\right)~,\\[0.2cm]
\psi^-_{s,x(\tau)}&\sim\de^{-r^2_+ s}\left(\begin{array}{c}
0\\
0 \\
\di \left(r_+^2+x^{12} +x^{34}\right) \\
x^{13}-x^{24}+\di (x^{23}+x^{14})
\end{array}
\right)~.
\end{aligned}
\end{equation}
Formulas \eqref{eq:MFieldDefinitions} yield the following Higgs field and gauge potential:
\begin{equation}\label{eq:Phik1}
\Phi=\begin{pmatrix}
\frac{\di}{2 r_-^{2}}&0\\
0&\frac{\di}{2 r_+^{2}}
\end{pmatrix}
\eand
A(\sigma)=\begin{pmatrix}
A^+(\sigma)&0\\
0&A^-(\sigma)
\end{pmatrix}~,
\end{equation}
where
\begin{equation}
A^+(\sigma)=\frac{\di}{2r_-^2(r_-^2+(x^{12}-x^{34}))}\left(\begin{array}{cc}
\xd^{3}(\sigma)(x^{23}-x^{14})+\xd^{4}(\sigma)(x^{13}+x^{24})\\
\xd^{4}(\sigma)(x^{23}-x^{14})-\xd^{3}(\sigma)(x^{13}+x^{24})\\
\xd^{1}(\sigma)(x^{14}-x^{23})+\xd^{2}(\sigma)(x^{13}+x^{24})\\
\xd^{2}(\sigma)(x^{14}-x^{23})-\xd^{1}(\sigma)(x^{13}+x^{24})
\end{array}
\right)~,
\end{equation}
and $A^-$ is obtained from $A^+$ by substituting $x^4(\sigma)\rightarrow -x^4(\sigma)$.

For $n=2$, we start from the solution $X^\mu$ to the Basu-Harvey equation with $\CA=A_4$ based on factorization:
\begin{equation}
 X^\mu=\frac{\tau_\mu}{\sqrt{2s}}~.
\end{equation}
The computation of the zero modes and the corresponding fields is a little more complicated in this case\cite{Palmer:2011vx}, but eventually one finds
\begin{equation}
\Phi=\begin{pmatrix}
\frac{\di}{r_-^{2}}&0\\
0&\frac{\di}{r_+^{2}}
\end{pmatrix}~.
\end{equation}
As expected, this is twice the Higgs field of \eqref{eq:Phik1}.

As a final consistency check, let us consider the reduction of the M-theory construction to the ADHMN construction with the Dirac operator $\nablas_{s,x}^{\rm IIA}$. The loops of $\CL \FR^4$ should be aligned along the M-theory direction $x^4$, which is compactified to a circle of radius $R=g_{\rm YM}$. That is, we restrict ourselves to loops of the form
\begin{equation}
 x^\mu(\tau)=x_0^\mu+g_{\rm YM}\delta_4^\mu\tau~~~\Rightarrow~~~\xd^\mu=g_{\rm YM}\delta_4^\mu~.
\end{equation}
In the solutions to the Basu-Harvey equation, we follow the standard Higgs mechanism for the reduction of M2-brane models to super Yang-Mills theory\cite{Mukhi:2008ux}. That is, we assume that the scalar field $X^4$ develops a vacuum expectation value in a 3-algebra direction: $\langle X^4\rangle=g_{\rm YM}\tau^4$. We can now expand all ingredients in our construction to highest order in $g_{\rm YM}$, and we find the corresponding objects in the ADHMN construction in the type IIA string theory interpretation. In particular, we have:
\begin{equation*}
 \begin{aligned}
x^{\mu\nu}:=\tfrac{1}{2}\oint \dd \tau\, \gamma^{\mu\nu}&x^\mu(\tau)\xd^\nu(\sigma) ~~~\rightarrow~~~g_{\rm YM}\gamma^{i4}x_0^i~,\\
  \dder{s}X^\mu=\tfrac{1}{3!}\eps^{\mu\nu\kappa\lambda}[X^\nu,X^\kappa,X^\lambda] &~~~\rightarrow~~~\dder{s}X^i=\frac{g_{\rm YM}}{2}\eps^{ijk}[X^j,X^k]+\CO(g_{\rm YM}^0)~,\\
  \nablas^{\rm M}_{s,x}~~~\rightarrow~~~\nablas_{s,x}^{\rm IIA}=-\gamma_5&\dder{s}+g_{\rm YM}\gamma^4\gamma^i(X^i-\di x^i)+\CO(g_{\rm YM}^0)~,
 \end{aligned}
\end{equation*}
where the fields $X^i$ take values in the Lie algebra $\asu(2)$, which is generated by $D(X^i,\tau_4)$. Note that the generators $\gamma^{\mu\nu}$ of $\sSpin(4)$ are reduced to $\gamma^{i4}$, which generate $\sSU(2)\cong \sSpin(3)\subset \sSpin(4)$. The ultra-local part of the self-dual string equation on loop space \eqref{eq:SDSequation} reduces directly to the Bogomolny equation $F=\star\dd\phi$.

\section{The construction for real and hermitian 3-algebras}

As mentioned above, the only finite-dimensional example of a 3-Lie algebra with all physically desirable properties is $A_4$. This 3-Lie algebra is conjectured to yield the effective description of two M2-branes. It is therefore necessary to extend our construction, in particular to the case of hermitian 3-algebras. The extension to real 3-algebras is almost trivial: We merely assume that the solution $X^\mu$ to the Basu-Harvey equation lives in a real 3-algebra, and the map $D:\CA\wedge \CA\rightarrow \frg_\CA$ is again the map from two elements of the real 3-algebra to its associated Lie algebra of inner derivations.

Physically more interesting and slightly more involved is the extension to hermitian 3-algebras. Here, we convert the four real scalar fields $X^\mu$ describing the transverse fluctuations of the M2-branes parallel to the worldvolume of the M5-branes into two complex fields
\begin{equation}
 Z^1:=X^1+\di X^2\eand Z^2:=-X^3-\di X^4~.
\end{equation}
Correspondingly, we introduce complex coordinates $z^1=x^1+\di x^2$ and $z^2=-x^3-\di x^4$ on $\FC^2\cong \FR^4$. For simplicity, we restrict ourselves to the hermitian 3-algebra $\CA$ given by $n\times n$ matrices as a vector space\footnote{For this 3-algebra, one has to impose a rather cumbersome reality condition: Traceless matrices are hermitian, while matrices proportional to the identity are antihermitian.}. The 3-bracket and the inner product on $\CA$ read as
\begin{equation}
 [a,b;c]:=a\bar c b-b \bar c a\eand(a,b):=\tr(\bar a b)~,~~~a,b,c\in\CA~,
\end{equation}
where $\bar{a}:=a^\dagger$. The ABJM model\cite{Aharony:2008ug} is now simply the BLG model generalized to fields taking values in the hermitian 3-algebra $\CA$.\cite{Bagger:2008se} The BPS equation in the ABJM model that corresponds to the Basu-Harvey equation is the following \cite{Gomis:2008vc,Terashima:2008sy,Hanaki:2008cu}:
\begin{equation}\label{eq:HermitianBasuHarvey}
 \dder{s}Z^\alpha=\tfrac{1}{2}(Z^\alpha\Zb_\beta Z^\beta-Z^\beta \Zb_\beta Z^\alpha)~,~~~\alpha,\beta=1,2~.
\end{equation}
Written in abstract 3-bracket notation with $D(a,b)\acton c:=[c,a;b]$, we have
\begin{equation}\label{eq:h3Basu-Harvey}
\dder{s}Z^\alpha=\tfrac{1}{2}[Z^\alpha,Z^\beta;Z^\beta]=-\tfrac{\di}{2}D(\di Z^\beta,Z^\beta)\acton Z^\alpha~.
\end{equation}
As in the case of real 3-algebras, the Dirac operator splits into two components
\begin{equation}
 \nablas_{s,z}:=\left(\begin{array}{cc}
\nablas^+_{s,z} & 0 \\
0 & \nablas^-_{s,z}
\end{array}\right)~,
\end{equation}
where e.g.\ $\nablas^+_{s,z}$ is constructed from a solution $Z^\alpha$ of \eqref{eq:HermitianBasuHarvey} as follows:
\begin{equation}
 \nablas^+_{s,z}=-\dder{s}-\tfrac{\di}{4}\sigma^{\mu\nu}\sigma^{\mu\nu}{}_\alpha{}^\beta\left(D(\di Z^\alpha,Z^\beta)-\oint\dd \tau\,z^{\alpha}(\tau)  \dot{\zb}_\beta(\tau) - \dot z^\alpha(\tau) \bar z_\beta(\tau)\right)~.
\end{equation}
Starting from the solutions to \eqref{eq:HermitianBasuHarvey} obtained from the product ansatz $Z^\alpha=r(s)G^\alpha$, where $G^\alpha$ is an element of the hermitian 3-algebra of $n\times n$-matrices, we obtain zero modes of the corresponding Dirac operator leading to the Higgs fields
\begin{equation}
 \Phi=\frac{\di n}{2r^2}\,\unit_2~.
\end{equation}
This field describes a self-dual string of charge $n$ and extends the examples for $n=1$ and $n=2$ based on 3-Lie algebras. Upon restricting loops and applying the Higgs mechanism, this construction, too, reduces to the ADHMN construction of monopoles with Dirac operator $\nablas_{s,x}^{\rm IIA}$.

\section{An M5-brane model on loop space}

Above we have demonstrated that the description of the BPS subsector of the M5-brane theory that is given by self-dual strings in terms of loop space works rather well. Let us now try to apply loop space to a description of the full model. Various models for an effective description of M5-branes have been proposed over the last years. Particularly interesting for our purposes is a proposal\cite{Lambert:2010wm} that assumes that all fields in the tensor multiplet assume values in a 3-Lie algebra $\CA$. Adding a gauge potential $A_\mu$ of a connection $\nabla$ as well as a 3-algebra valued vector $C^\mu$, the field equations
\begin{equation}\label{eq:eomTensor}
\begin{aligned}
 H_{\mu\nu\kappa}-\tfrac{1}{3!}\eps_{\mu\nu\kappa\lambda\rho\sigma}H^{\lambda\rho\sigma}&=0~,\\
 \nabla^2 X^I-\tfrac{\di}{2}[\bar{\Psi},\Gamma_\nu\Gamma^I\Psi,C^\nu]+[X^J,C^\nu,[X^J,C_\nu,X^I]]&=0~,\\
 \Gamma^\mu\nabla_\mu\Psi-[X^I,C^\nu,\Gamma_\nu\Gamma^I\Psi]&=0~,\\
 \nabla_{[\mu}H_{\nu\kappa\lambda]}+\tfrac{1}{4}\eps_{\mu\nu\kappa\lambda\sigma\tau}[X^I,\nabla^\tau X^I,C^\sigma]+\tfrac{\di}{8}\eps_{\mu\nu\kappa\lambda\sigma\tau}[\bar{\Psi},\Gamma^\tau\Psi,C^\sigma]&=0~,\\
F_{\mu\nu}-D(C^\lambda,H_{\mu\nu\lambda})&=0~,\\
\nabla_\mu C^\nu=D(C^\mu,C^\nu)&=0~,\\
D(C^\rho,\nabla_\rho X^I)=D(C^\rho,\nabla_\rho\Psi)=D(C^\rho,\nabla_\rho H_{\mu\nu\lambda})&=0~.
\end{aligned}
\end{equation}
are invariant under the supersymmetry transformations
\begin{equation}
 \begin{aligned}
  \delta X^I&=\di \epsb \Gamma^I\Psi~,\\
  \delta \Psi&=\Gamma^\mu\Gamma^I\nabla_\mu X^I\eps+\tfrac{1}{2\times 3!}\Gamma_{\mu\nu\lambda}H^{\mu\nu\lambda}\eps-\tfrac{1}{2}\Gamma^{IJ}\Gamma_\lambda[X^I,X^J,C^\lambda]\eps~,\\
  \delta H_{\mu\nu\lambda}&=3\di \epsb\Gamma_{[\mu\nu}\nabla_{\lambda]}\Psi+\di\epsb\Gamma^I\Gamma_{\mu\nu\lambda\kappa}[X^I,\Psi,C^\kappa]~,\\
  \delta A_\mu&=\di\epsb\Gamma_{\mu\lambda} D(C^\lambda,\Psi)~,\\
  \delta C^\mu&=0~.
 \end{aligned}
\end{equation}

The field equation $D(C^\mu,C^\nu)=0$ implies a factorization of the vector $C^\mu$ into $c^\mu C$, where $C$ is an element of the 3-Lie algebra $\CA$ and $c^\mu$ is a vector in $\FR^{1,5}$. The appearance of this additional vector $c^\mu$ on spacetime suggests to identify it with the M-theory direction. Going to loop space, we correspondingly identify $c^\mu$ with the tangent $\xd^\mu(\tau)$ to the loop $x^\mu(\tau)$. In this picture, we obtain a very natural interpretation for the equation $F_{\mu\nu}-D(C^\lambda,H_{\mu\nu\lambda})=0$ in \eqref{eq:eomTensor}:
\begin{equation}\label{eq:transgressionCondition}
 \cF_{\mu\nu}(x)=D\big(C^\lambda,H_{\mu\nu\lambda}(x(\tau))\big)=D\big(C,H_{\mu\nu\lambda}(x(\tau))\,\xd^\lambda(\tau)\big)~,
\end{equation}
which is a non-abelian form of the transgression map \eqref{eq:Transgression}. Here, we added a $\circ$ to indicate that the field $\cF$ lives on loop space. Introducing moreover the fields
\begin{equation}\label{eq:3LAtransgression}
 \cX^I(x(\tau)):=R~D(C,X^I(x(\tau)))\eand\cPsi(x(\tau)):=\Gamma^\rho \xd_\rho D(C,\Psi(x(\tau)))~,
\end{equation}
the field equations
\begin{equation}\label{eq:eomsummary}
\begin{aligned}
 \nabla^2 \cX^I+\tfrac{\di}{2}\tfrac{1}{R}\xd^\nu[\bar{\cPsi},\Gamma_\nu\Gamma^I\cPsi]+ [\cX^J,[\cX^J,\cX^I]]&=0~,\\
\tfrac{1}{R} \Gamma^{\mu\nu}\xd_\nu\nabla_\mu\cPsi-\Gamma^I [\cX^I,\cPsi]&=0~,\\
\nabla_\mu \cF^{\mu\nu}+[\cX^I,\nabla^\nu \cX^I]+\di\left( [\bar{\cPsi},\Gamma^\nu\cPsi] - \tfrac{1}{R^2}\xd^\sigma \xd^\nu[\bar \cPsi, \Gamma_\sigma \cPsi]\right)&=0
\end{aligned}
\end{equation}
are invariant under the supersymmetry transformations
\begin{equation}\label{eq:susysummary}
\begin{aligned}
 \delta \cX^I&=\tfrac{1}{R}\di \epsb \Gamma^I\xd^\rho\Gamma_\rho\cPsi~,\\
 \delta \cA_{\mu}&=\tfrac{1}{R^2} \di\epsb \Gamma_{\mu\lambda}\Gamma_\rho\xd^\lambda\xd^\rho\cPsi~,\\
  \delta \cPsi&=\tfrac{1}{R}\Gamma^{\nu\mu}\xd_\nu\Gamma^I\nabla_\mu\cX^I\eps+\tfrac{1}{2}\Gamma_{\mu\nu}\cF^{\mu\nu}\eps-\tfrac{1}{2}\Gamma^{IJ}[\cX^I,\cX^J]\eps~.
\end{aligned}
\end{equation}
Note that all the above equations are local on the loop $x^\mu(\tau)$.

There are, however, a few open questions concerning this rewriting: The additional conditions
\begin{equation}\label{eq:constraintssummary}
\xd^\mu \nabla_\mu \cX^I = \xd^\mu \nabla_\mu \cPsi  = \xd^\mu \nabla_\mu \cF_{\nu\lambda} =0~.
\end{equation}
should merely state that the fields are invariant under reparametrizations. The covariant derivatives $\nabla_\mu$, however, acts only on the zero-modes of the fields:
\begin{equation}
 \nabla_\mu:=\oint\dd \tau\,\frac{\delta}{\delta x^\mu(\tau)}+A_{\mu}(\tau)\acton~,
\end{equation}
and this should be extended to a full loop space derivative. At the moment, it is not clear how to do this while preserving supersymmetry.

To compare this model with our generalization of the ADHMN construction, we can derive the BPS equations from the supersymmetry transformations on loop space \eqref{eq:susysummary}. This yields indeed the ultra-local part of the loop space self-dual string equation \eqref{eq:SDSequation}.

\section{Summary and recent developments}

The ADHMN-like construction we reviewed above arose very naturally from the assumption that self-dual strings should be described in terms of loop space: The necessary lift of the Dirac operator of the original construction as well as the non-abelian generalization of the equation were readily derived. Also, the Basu-Harvey equation features in the construction exactly in the expected way and takes over the role the Nahm equation played in the original ADHMN construction.

Explicit examples of solutions to the loop space self-dual string equation can be constructed. Moreover, the reduction to the ADHMN construction of monopoles is rather obvious. Our construction extends to the more general real and hermitian 3-algebras, and it is therefore compatible with the ABJM model. Also, we obtained naturally a gauge group structure $G\times G$, which fits well with a recent proposals for M5-brane equations\cite{Chu:2011fd}.

Besides the na\"ive argument for the appearance of loop space given in the introduction, there is further evidence that this approach is appropriate. First of all, the notion of a monopole bag together with the associated Nahm construction have been lifted to M-theory\cite{Harland:2012cj}. Recall that a monopole bag is a double-scaling limit of infinitely many monopoles forming a surface diffeomorphic to a sphere, where the ratio of the monopole number and the Yang-Mills coupling constant is kept fixed. In this double-scaling limit, all equations become abelian, and the description of self-dual string bags is therefore known. It was checked that the infinite-charge Nahm construction has a loop space version, which produces solutions to an abelian form of the loop space self-dual string equation\cite{Harland:2012cj}.

Further motivation for the use of loop space also comes from a different perspective. One might wonder about the interpretation of the fuzzy funnel in the case of M2-branes ending on M5-branes. Recall that in the $\sSO(3)$-invariant situation of D1-branes ending on D3-branes, the points in the worldvolume of the D1-branes polarize into fuzzy two-spheres. Analogously one would expect that points on M2-branes polarize into fuzzy three-spheres. A satisfyingly precise notion of fuzzy 3-spheres is so far unknown. Roughly speaking, the prequantum line bundle over $S^2$ should be replaced by a prequantum gerbe over $S^3$. The Hilbert space of the quantization, which in the case of $S^2$ is given by global holomorphic sections of the prequantum line bundle, should here be obtained from global holomorphic sections of the prequantum gerbe. While the latter notion seems rather opaque, one can transgress the problem to loop space: The loop space $\CL S^3$ of $S^3$ carries the structure of a K\"ahler manifold\cite{0817647309}, and the prequantum gerbe over $S^3$ is transgressed to a prequantum line bundle over $\CL S^3$. One can then quantized (a subset of the) functions on the space $\CL S^3$ by identifying them with linear operators on the space of global holomorphic sections of the transgressed abelian gerbe over $S^3$. Such an approach to quantization has been suggested previously\cite{Samann:2011zq,Saemann:2011yi,Saemann:2012ex}, where it was used to quantize the loop space of $\FR^3$. It could be shown that the quantization of the loop space reduces as expected to the Moyal quantization of $\FR^2$ upon imposing the usual restrictions reducing M-theory to string theory.

The fact that an ADHMN-like construction for self-dual strings exists at all is by now no longer surprising. Underlying both the original ADHM and ADHMN constructions are corresponding twistor geometric interpretations of the instanton and monopole equations. Here, a Penrose-Ward transform establishes a bijection between certain holomorphic bundles over a twistor space and solutions to the relevant equations. Quite recently, a twistor space was found\cite{Saemann:2011nb}, that yields a bijection between certain holomorphic gerbes on this twistor space and (abelian) solutions to the self-dual string equation. In principle it is possible to transgress this picture to loop space and to obtain the twistor picture underlying our loop space self-dual string equation.

Moreover, the twistor construction using abelian gerbes\cite{Saemann:2011nb,Mason:2011nw} has recently been extended to the non-abelian case\cite{Saemann:2012uq}. We therefore know that a full ADHMN-like construction of self-dual strings based on ordinary space and non-abelian gerbes has to exist.

The existence of our construction leads to another question. Recall that the ADHMN construction is a special case of a Nahm transform mapping instanton solutions on a four-torus to instanton solutions on the dual four-torus. It is therefore a map between solutions to the same theory on different spaces and in the special case of the ADHMN construction, one has a map between solutions to dimensional reductions of the instanton equation. Therefore one would also expect the M-theory lift to be a map between solutions to closely related theories. The Basu-Harvey equation and the self-dual string equation, however, appear to be very different. A first step towards resolving this issue is the observation that all 3-Lie algebras (as well as the generalized real and hermitian 3-algebras) are special cases of differential crossed modules or strict Lie 2-algebras\cite{Palmer:2012ya}. These strict Lie 2-algebras, in turn, are structure groups of categorified principal bundles, which contain in particular non-abelian gerbes as special cases. Assuming that the gauge structure of the M5-brane model yet to be found will be defined by a non-abelian gerbe, we can conclude that at least the gauge structure of both M2- and M5-brane models do agree. Moreover, in the M2-brane model the gauge two-form curvature carries no degrees of freedom, as it is fixed via the equations of motion by the matter fields. Something analogous happens when working with non-abelian gerbes: Here, a gauge two-form curvature is fixed by the two-form potential via the so-called fake curvature condition and just as in Chern-Simons matter theories, the gauge field does not carry any additional degrees of freedom. Altogether, it seems that M2- and M5-brane models might not be as different as it appeared initially. This, together with the fact that more and more M5-brane models are being proposed\cite{Samtleben:2011fj,Chu:2011fd,Chu:2012um}, gives reason for careful optimism about the existence of an effective description of M5-branes.

\section*{Acknowledgments}

I would like to thank Sam Palmer and Costis Papageorgakis for very pleasant collaborations on the papers\cite{Palmer:2011vx,Papageorgakis:2011xg} that I reviewed here. Moreover, I am grateful to Sergey Cherkis for the invitation to Berkeley as well as for many inspiring and entertaining discussions.  I would also like to thank Martin Wolf for comments on a draft of this review. This work was supported by a Career Acceleration Fellowship from the UK Engineering and Physical Sciences Research Council.



\begin{thebibliography}{10}

\bibitem{Basu:2004ed}
A.~Basu and J.~A. Harvey, {\em Nucl. Phys. B}  {\bf 713}, 136 (2005).

\bibitem{Bagger:2007jr}
J.~Bagger and N.~D. Lambert, {\em Phys. Rev. D} {\bf 77}, 065008 (2008).

\bibitem{Gustavsson:2007vu}
A.~Gustavsson, {\em Nucl. Phys. B} {\bf 811}, 66 (2009).

\bibitem{Filippov:1985aa}
V.~T. Filippov, {\em Sib. Mat. Zh.} {\bf 26}, 126 (1985).

\bibitem{Aharony:2008ug}
O.~Aharony, O.~Bergman, D.~L. Jafferis and J.~M. Maldacena, {\em JHEP} {\bf
  0810}, 091 (2008).

\bibitem{Bagger:2008se}
J.~Bagger and N.~Lambert, {\em Phys. Rev. D} {\bf 79}, 025002 (2009).

\bibitem{Bagger:2012jb}
J.~Bagger, N.~Lambert, S.~Mukhi and C.~Papageorgakis, preprint, arXiv:1203.3546, 2012.

\bibitem{Drukker:2010nc}
N.~Drukker, M.~Marino and P.~Putrov, {\em Commun. Math. Phys.} {\bf 306}, 511
  (2011).

\bibitem{Witten:2007ct}
E.~Witten, preprint, arXiv:0712.0157, 2007.

\bibitem{Chu:2012um}
C.-S. Chu and S.-L. Ko, {\em JHEP} {\bf 1205}, 028 (2012).

\bibitem{Saemann:2010cp}
C.~Saemann, {\em Commun. Math. Phys.} {\bf 305}, 513 (2011).

\bibitem{Palmer:2011vx}
S.~Palmer and C.~Saemann, {\em JHEP} {\bf 1110}, 008 (2011).

\bibitem{Schreiber:2005ff}
U.~Schreiber, preprint, arXiv:hep-th/0509163, 2005.

\bibitem{Gustavsson:2005aq}
A.~Gustavsson, {\em JHEP} {\bf 0601}, 165 (2006).

\bibitem{Gustavsson:2006ie}
A.~Gustavsson, {\em JHEP} {\bf 0612}, 066 (2006).

\bibitem{Gustavsson:2008dy}
A.~Gustavsson, {\em JHEP} {\bf 0804}, 083 (2008).

\bibitem{Diaconescu:1996rk}
D.-E. Diaconescu, {\em Nucl. Phys.} {\bf B503}, 220 (1997).

\bibitem{Tsimpis:1998zh}
D.~Tsimpis, {\em Phys. Lett. B} {\bf 433}, 287 (1998).

\bibitem{Nahm:1981nb}
W.~Nahm, Talk presented at Int. Summer Inst. on Theoretical Physics, Freiburg, West Germany, Aug 31 - Sep 11, 1981.

\bibitem{Hitchin:1983ay}
N.~J. Hitchin, {\em Commun. Math. Phys.} {\bf 89}, 145.

\bibitem{Papageorgakis:2011xg}
C.~Papageorgakis and C.~Saemann, {\em JHEP} {\bf 1105}, 099 (2011).

\bibitem{Lambert:2010wm}
N.~Lambert and C.~Papageorgakis, {\em JHEP} {\bf 1008}, 083 (2010).

\bibitem{Howe:1997ue}
P.~S. Howe, N.~D. Lambert and P.~C. West, {\em Nucl. Phys. B} {\bf 515}, 203 (1998).

\bibitem{Myers:1999ps}
R.~C. Myers, {\em JHEP} {\bf 12}, 022 (1999).

\bibitem{Cherkis:2008qr}
S.~Cherkis and C.~Saemann, {\em Phys. Rev. D} {\bf 78}, 066019 (2008).

\bibitem{deAzcarraga:2010mr}
J.~A. de~Azcarraga and J.~M. Izquierdo, {\em J. Phys.} {\bf A43}, 293001
  (2010).

\bibitem{Murray:2007ps}
M.~K. Murray (2007), in: {\em The Many Facets of Geometry: A Tribute to Nigel
  Hitchin,} eds.\ O.\ Garcia-Prada {\em et al.} (Oxford
  University Press, 2010).

\bibitem{Hitchin:2010qz}
N.~Hitchin (2010), preprint, arXiv:1008.0973, 2010.

\bibitem{Chatterjee:1998}
D.~S. Chatterjee, Ph.D. thesis, Trinity College, Cambridge, 1998.

\bibitem{0817647309}
J.-L. Brylinski, {\em Loop spaces, characteristic classes and geometric quantization} (Birkh{\"a}user, 2007).

\bibitem{Schenk:1986xe}
H.~Schenk, {\em Commun. Math. Phys.} {\bf 116}, 177.

\bibitem{Campos:2000de}
V.~L. Campos, G.~Ferretti and P.~Salomonson, {\em JHEP} {\bf 0012}, 011 (2000).

\bibitem{Mukhi:2008ux}
S.~Mukhi and C.~Papageorgakis, {\em JHEP} {\bf 0805}, 085 (2008).

\bibitem{Gomis:2008vc}
J.~Gomis, D.~Rodriguez-Gomez, M.~Van~Raamsdonk and H.~Verlinde, {\em JHEP} {\bf
  0809}, 113 (2008).

\bibitem{Terashima:2008sy}
S.~Terashima, {\em JHEP} {\bf 0808}, 080 (2008).

\bibitem{Hanaki:2008cu}
K.~Hanaki and H.~Lin, {\em JHEP} {\bf 0809}, 067 (2008).

\bibitem{Chu:2011fd}
C.-S. Chu, preprint, arXiv:1108.5131, 2011.

\bibitem{Harland:2012cj}
D.~Harland, S.~Palmer and C.~Saemann, preprint, arXiv:1204.6685, 2012.

\bibitem{Samann:2011zq}
C.~Saemann and R.~J. Szabo, {\em PoS} {\bf CNCFG2010}, 005 (2010).

\bibitem{Saemann:2011yi}
C.~Saemann and R.~J. Szabo, Proc. of the 4-th Annual Meeting of the
  European Non-Commutative Geometry Network, Bucharest, 25.4.-30.4. 2011.

\bibitem{Saemann:2012ex}
C.~Saemann and R.~J. Szabo, preprint, arXiv:1203.5921, 2012.

\bibitem{Saemann:2011nb}
C.~Saemann and M.~Wolf, preprint, arXiv:1111.2539, 2011.

\bibitem{Mason:2011nw}
L.~Mason, R.~Reid-Edwards and A.~Taghavi-Chabert, preprint, arXiv:1111.2585, 2011.

\bibitem{Saemann:2012uq}
C.~Saemann and M.~Wolf (2012), preprint, arXiv:1205.3108, 2012.

\bibitem{Palmer:2012ya}
S.~Palmer and C.~Saemann (2012), preprint, arXiv:1203.5757, 2012.

\bibitem{Samtleben:2011fj}
H.~Samtleben, E.~Sezgin and R.~Wimmer, {\em JHEP} {\bf 1112}, 062 (2011).

\end{thebibliography}

\end{document}